\documentclass{iopart}

\usepackage{iopams}
\usepackage{graphicx}

\usepackage[draft]{hyperref}

\usepackage[expproduct=cdot,noload=abbr,alsoload=hep]{siunitx}

\usepackage[caption=false, font=footnotesize, justification=RaggedRight]{subfig}

\usepackage{paralist}

\graphicspath{{./}}
\DeclareGraphicsExtensions{.eps}

\newunit{\nuclearNumber}{A}
\newunit{\AGeV}{\nuclearNumber\giga\electronvolt}
\newunit{\GeV}{\giga\electronvolt}
\newunit{\TeV}{\tera\electronvolt}
\newunit{\ATeV}{\nuclearNumber\tera\electronvolt}
\newunit{\fmc}{\femto\metre\per\clight}
\newunit{\fm}{\femto\metre}

\newcommand{\AuAu}{Au\,+\,Au}
\newcommand{\PbPb}{Pb\,+\,Pb}
\newcommand{\ProtonProton}{p\,+\,p}

\newcommand{\vect}[1]{\boldsymbol{\mathbf{#1}}}

\begin{document}
\title[Jets and $v_{2}$ in Partonic Transport]{Jet quenching and elliptic flow at RHIC and LHC within a pQCD-based partonic transport model}

\author{O Fochler$^1$, J Uphoff$^1$, Z Xu$^{1,2}$ and C Greiner$^1$}

\address{$^1$ Institut f\"ur Theoretische Physik, Goethe-Universit\"at Frankfurt am Main\\
Max-von-Laue-Stra\ss{}e~1, D-60438 Frankfurt am Main, Germany}

\address{$^2$ Frankfurt Institute for Advanced Studies (FIAS)\\
  Ruth-Moufang-Stra\ss{}e~1, D-60438 Frankfurt am Main, Germany}

\ead{fochler@th.physik.uni-frankfurt.de}

\begin{abstract}
Fully dynamic simulations of heavy ion collisions at RHIC and at LHC energies within the perturbative QCD-based partonic transport model BAMPS (Boltzmann Approach to Multi-Parton Scatterings) are presented, focusing on the simultaneous investigation of jet quenching and elliptic flow. The model features inelastic $2 \leftrightarrow 3$ processes based on the Gunion-Bertsch matrix element and has recently been extended to include light quark degrees of freedom, allowing for direct comparison to hadronic data on the nuclear modification factor via a fragmentation scheme for high-$p_{T}$ partons. The nuclear modification factor of neutral pions in central \AuAu{} collisions at RHIC energy is compared to experimental data. Furthermore first results on the nuclear modification factor and the integrated elliptic flow of charged hadrons in \PbPb{} collisions at LHC are presented and compared to recent ALICE data. These investigations are complemented by a study on the suppression of $D$-mesons at LHC based on elastic interactions with the medium.
\end{abstract}

\section{Introduction}

The suppression of high-$p_{T}$ spectra with respect to a scaled \ProtonProton{} reference, quantified in terms of the nuclear modification factor $R_{AA}$, and the strong collectivity of the medium, quantified in terms of the Fourier coefficient $v_{2}$, the elliptic flow, have been established by experiments at the \emph{Relativistic Heavy Ion Collider} (RHIC). First results from the recently commissioned \emph{Large Hadron Collider} (LHC) have confirmed these findings at an order of magnitude higher collision energy, showing remarkable quantitative similarities beyond a good qualitative agreement.
The suppression of particles with high transverse momentum is commonly attributed to an energy loss on the partonic level during the quark-gluon plasma (QGP) stage of the evolution of the medium. Comparison of $v_{2}$ measurements to hydrodynamic simulations have established that the elliptic flow builds up early and is thus also governed by the evolution of the QGP medium. It is notoriously difficult to combine both aspects of the QPG evolution---the suppression of rare, high-$p_{T}$ probes and the collective behavior of the bulk particles---into a common model. The work presented in this article explores the capabilities of microscopic partonic transport models in this respect.

\section{The transport model BAMPS}

The microscopic transport model BAMPS (\emph{Boltzmann Approach to Multi-Parton Scatterings}) \cite{Xu:2004mz} is applied to simulate the time evolution of the hot partonic medium that is created in heavy ion collisions at RHIC and LHC. It is based on matrix elements in leading order perturbative QCD (pQCD) and consistently features inelastic $2 \leftrightarrow 3$ interactions. Rapid thermalization of the simulated partonic medium on the order of \SI{1}{\fmc} has been found for \AuAu{} collisions at RHIC energies \cite{Xu:2004mz} and can be confirmed also for \PbPb{} collisions at LHC energies.

Originally limited to a purely gluonic medium ($N_{f}=0$) the model has been extended to include light quarks ($N_{f}=3$). These are considered to be massless Boltzmann particles as are the gluons. Binary interactions involving light quarks and gluons are computed from leading order pQCD cross sections in small angle approximation. Radiative and annihilation processes, $gg \leftrightarrow ggg$, are based on the Gunion-Bertsch matrix element \cite{Gunion:1981qs}
\begin{equation}
  \label{eq:GunionBertsch_matrix_element}
  \left| \mathcal{M}_{gg \rightarrow ggg} \right|^{2} = 
  \left( \frac{72 \pi^{2} \alpha_{s}^{2}\, s^{2}}{(\vect{q}_{\perp}^{2} + m_{D}^{2})^{2}} \right)
  \left( \frac{48 \pi \alpha_{s}\, \vect{q}_{\perp}^{2}}{\vect{k}_{\perp}^{2} \left[ (\vect{k}_{\perp} - \vect{q}_{\perp})^{2} + m_{D}^{2} \right]} \right)
  \text{.}
\end{equation}
$2 \leftrightarrow 3$ processes involving light quarks are also computed from \eref{eq:GunionBertsch_matrix_element} based on a factorization of the Gunion-Bertsch matrix element into a collisional part and a radiation probability $\left| \mathcal{M}_{\text{GB}} \right|^{2} = \left| \mathcal{M}_{\text{coll}} \right|^{2} \, P^{g}$. Applying the small angle approximation to $\left| \mathcal{M}_{\text{coll}} \right|^{2}$, the computation of $2 \leftrightarrow 3$ processes involving light quarks is reduced to a scaling of \eref{eq:GunionBertsch_matrix_element} by color and symmetry factors.

The cross sections and matrix elements are screened by a Debye mass $m_{D}^{2} = d_G \pi \alpha_{s} \int \frac{d^{3}p}{(2\pi)^{3}} \frac{1}{p} \left(N_{c}f_{g} + N_{f}f_{q} \right)$ that is dynamically computed from the current distributions of gluons and quarks. The Landau-Pomeranchuk-Migdal (LPM) effect is modeled via the introduction of a cutoff $\Theta \left( \lambda - \tau \right)$ into \eref{eq:GunionBertsch_matrix_element} that effectively discards coherent contributions from multiple induced gluon radiation. This is done by a comparison of the mean free path $\lambda$ to the formation time of the radiated gluon $\tau$. More details on the modeling of the LPM effect and on the consequences arising from this implementation can be found in \cite{Fochler:2010wn}.

\section{Nuclear modification factor and elliptic flow for gluons and light quarks}

As established in \cite{Xu:2007jv,Xu:2008av,Fochler:2008ts} the matter in BAMPS simulations of \AuAu{} collisions at RHIC energies exhibits a strong degree of collectivity with an integrated $v_{2}$ that is in good agreement with experimental results over a large centrality range for a fixed strong coupling of $\alpha_{s}=0.3$ and a freeze-out energy density $\varepsilon_{c} = \SI{0.6}{\GeV\per\femto\metre\tothe{3}}$. These parameters are used for all calculations that are presented in this section.

\begin{figure}[tbh]
  \sisetup{tophrase=dash}
  \centering
  \subfloat[Nuclear modification factor $R_{AA}$ for neutral pions from BAMPS simulations of \AuAu{} at \SI{200}{\AGeV} compared to PHENIX results \cite{Adare:2008qa} at \SIrange{0}{10}{\percent} centrality. $R_{AA}$ of gluons and quarks is shown for comparison. Lines indicate $R_{AA}$ computed from fits to the parton spectra, while symbols indicate $R_{AA}$ computed directly from the parton spectra as obtained from BAMPS.]{%
  \label{fig:Fig1:RAA_RHIC}
  \includegraphics[width=0.49\textwidth]{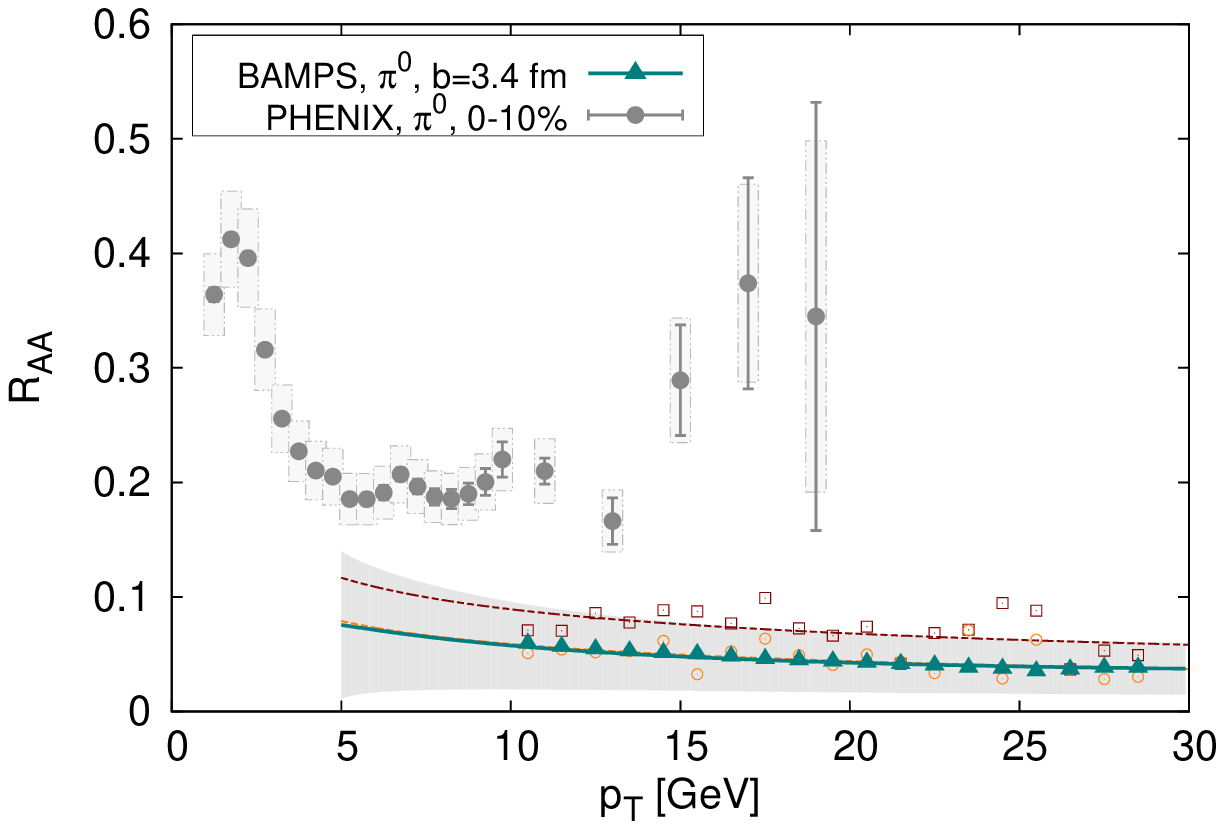}%
  }
  \subfloat[Nuclear modification factor $R_{AA}$ of charged hadrons, gluons and quarks from BAMPS simulations of \PbPb{} at $b=\SI{0}{\fm}$ compared to results from ALICE for \SIrange{0}{5}{\percent} central \PbPb{} collisions \cite{Aamodt:2010jd}. Lines and symbols as in \fref{fig:Fig1:RAA_RHIC}. For comparison the $R_{AA}$ of charged hadrons from simulations of \AuAu{} at \SI{200}{\AGeV} and $b=\SI{0}{\fm}$ is also shown.]{%
  \label{fig:Fig1:RAA_LHC}
  \includegraphics[width=0.49\textwidth]{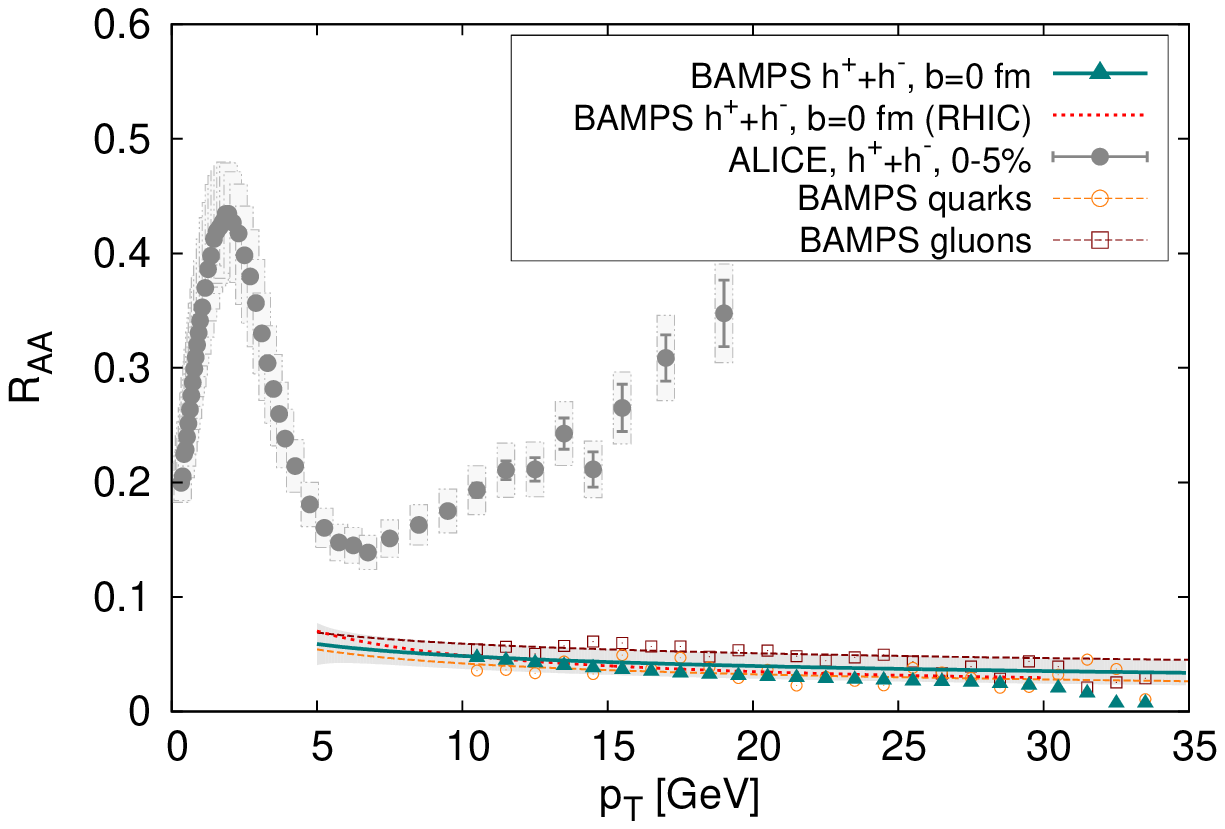}%
  }
  \caption{Nuclear modification factor at RHIC and LHC.}
\label{fig:Fig1}
\end{figure}

Using this setup and mini-jet initial conditions with $p_{0}=\SI{1.4}{\GeV}$, \fref{fig:Fig1:RAA_RHIC} shows the nuclear modification factor $R_{AA}$ obtained from BAMPS simulations of central, \SIrange[tophrase=dash]{0}{10}{\percent}, \AuAu{} collisions at \SI{200}{\AGeV}. The results are both shown on the partonic level for gluons and light quarks and on the hadronic level for neutral pions based on AKK fragmentation functions \cite{Albino:2008fy}. The suppression of high-$p_{T}$ particles in simulations with BAMPS is distinctly stronger than the experimentally observed suppression. This is due to 
\begin{inparaenum}[\itshape a\upshape)]
\item a strong energy loss that is caused by a complex interplay of the Gunion-Bertsch matrix element and the effective implementation of the LPM effect \cite{Fochler:2010wn},
\item a conversion of quark into gluon jets in $2 \rightarrow 3$ interactions and
\item a small difference in the energy loss of gluons and quarks caused by the iterative computation of interaction rates required by the inclusion of the LPM cutoff into \eref{eq:GunionBertsch_matrix_element}.
\end{inparaenum}

Going to the LHC energy of $\sqrt{s_{NN}}=\SI{2.76}{\TeV}$ and using \textsc{Pythia} initial conditions \cite{Uphoff:2010sh}, \fref{fig:Fig1:RAA_LHC} shows the nuclear modification factor of charged hadrons for central \PbPb{} collisions. The results are not significantly different from those at RHIC energy and again the suppression of high-$p_{T}$ particles is overestimated compared to the experimental data. Also the upward trend towards large $p_{T}$ present in the ALICE data is not reproduced.

The simulated differential $v_{2}$ of central and semi-central \PbPb{} collisions at \SI{2.76}{\ATeV} shows no significant deviation from the \AuAu{} results at \SI{200}{\AGeV} in the low to intermediate $p_{T}$ region which is in agreement with experimental findings \cite{Aamodt:2010jd}. Following from an increase in the mean transverse momenta, the integrated $v_{2}$ however is larger. \Fref{fig:Fig2:v2_LHC} shows the integrated $v_{2}$ for \PbPb{} at LHC as a function of centrality compared to experimental data from ALICE \cite{Aamodt:2010jd}. Up to roughly \SI{40}{\percent} centrality the agreement is very good, going to more peripheral collisions the simulated $v_{2}$ exhibits a drop that is distinctly more pronounced than in the experimental data. The cause of this rapid drop is currently under systematic investigation, comparing for example different initial conditions and freeze-out prescriptions.

\section{Heavy quarks}

\begin{figure}[tbh]
  \sisetup{tophrase=dash}
  \centering
  \subfloat[Integrated $v_{2}$ of gluons and light quarks ($|y|<0.8$) as a function of centrality for \PbPb{} at \SI{2.76}{\ATeV} compared to the measured $v_{2}$ of charged particles from ALICE \cite{Aamodt:2010pa}.]{%
  \label{fig:Fig2:v2_LHC}
  \includegraphics[width=0.48\textwidth]{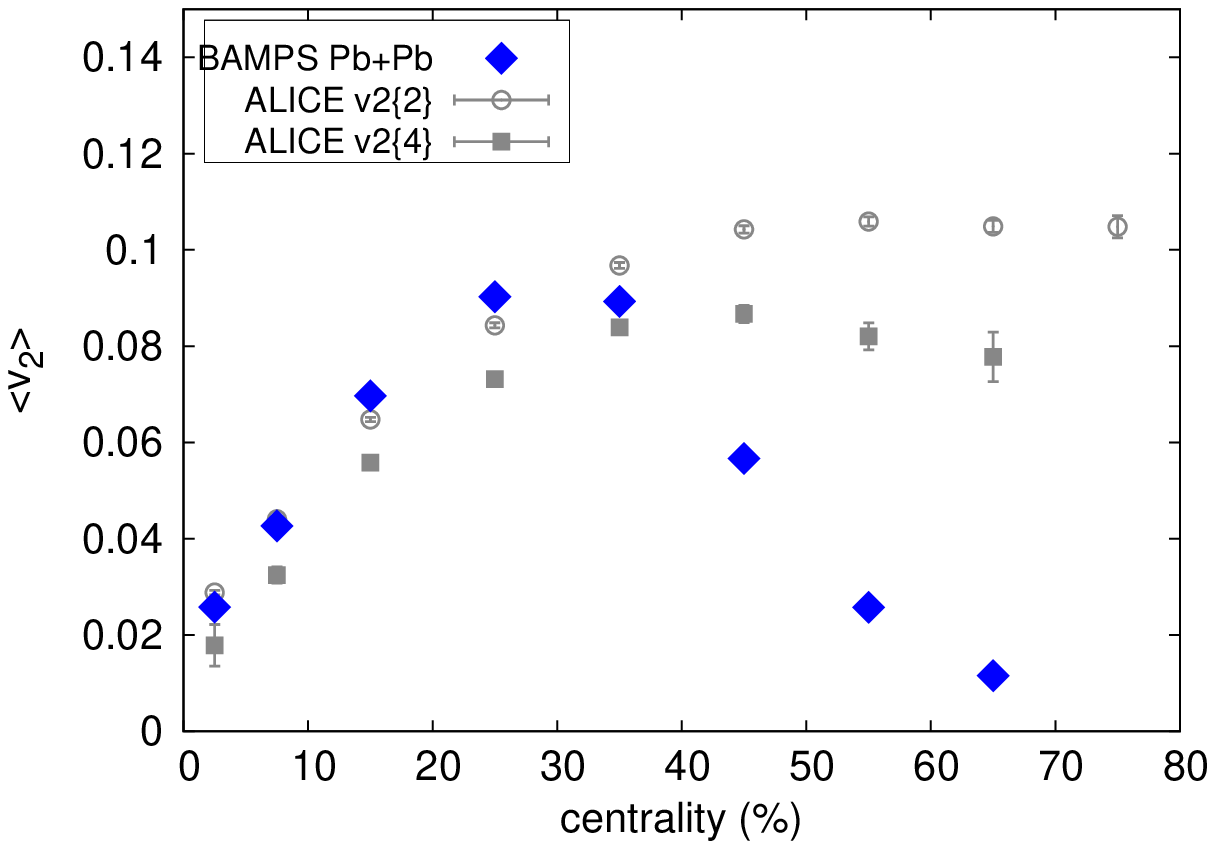}%
  }%
  \hspace{0.1em}%
  \subfloat[Nuclear modification factor $R_{AA}$ for $D$-mesons ($|y|<0.5$) for \PbPb{} collisions at \SI{2.76}{\ATeV} and $b=\SI{4.5}{\fm}$ for the same setup as in \cite{Uphoff:2011ad}. For comparison preliminary data from ALICE is shown for \SIrange{0}{20}{\percent} centrality \cite{alice_dmeson}.]{%
  \label{fig:Fig2:Dmesons_LHC}
  \includegraphics[width=0.48\textwidth]{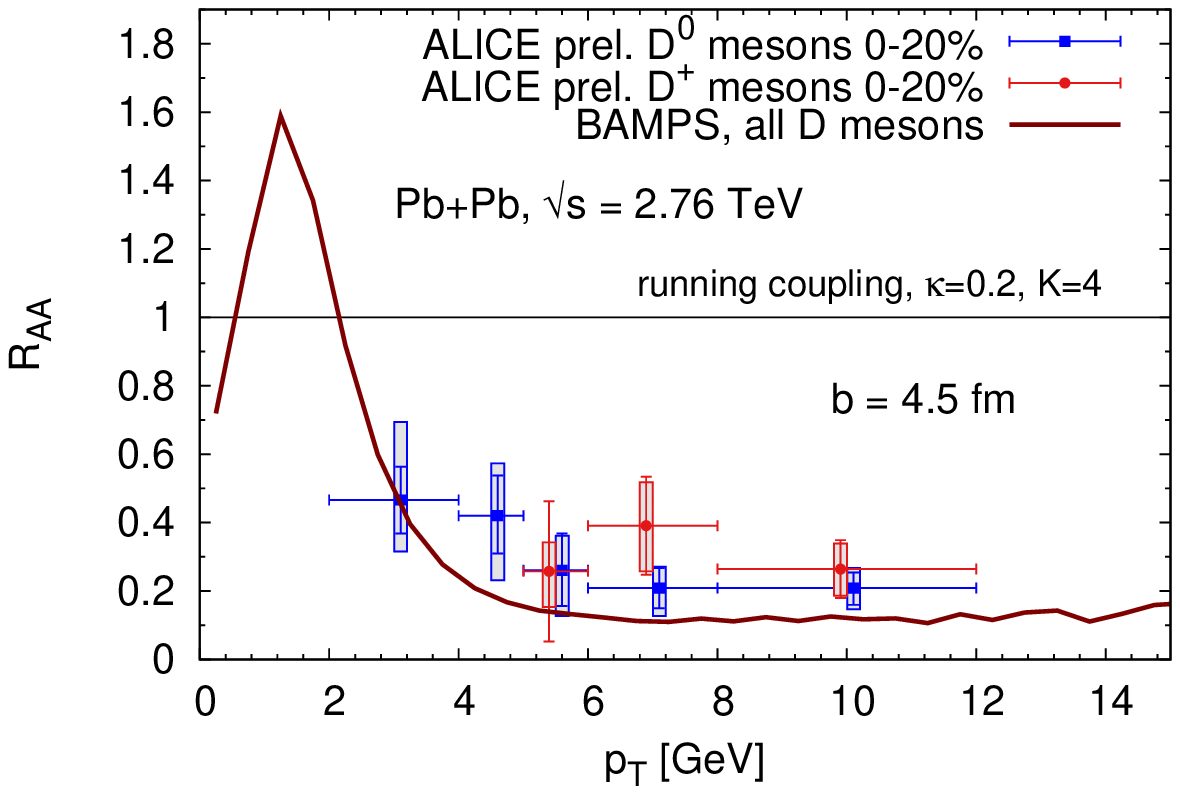}%
  }
  \caption{Integrated $v_{2}$ and $D$-meson $R_{AA}$ at LHC.}
\label{fig:Fig2}
\end{figure}

The BAMPS framework is also extensively applied to the investigation of heavy quark phenomena \cite{Uphoff:2010sh,Uphoff:2010bv,Uphoff:2011ad}. \Fref{fig:Fig2:Dmesons_LHC} compares the $R_{AA}$ of $D$-mesons obtained from BAMPS to preliminary ALICE data \cite{alice_dmeson}. Heavy quark elastic interactions with the gluonic medium ($N_{f}=0$) are implemented with running coupling and a Debye screening motivated from HTL calculations \cite{Uphoff:2011ad}. To account for radiative contributions the cross section is multiplied by a factor $K=4$ which gives a very good agreement with experimental data for $R_{AA}$ and $v_2$ of heavy flavor electrons at RHIC \cite{Uphoff:2011ad}. It will be checked in a forthcoming study whether the implementation of radiative heavy quark processes has indeed the same effect as scaling the elastic cross section with a constant $K$ factor.

\section{Summary}

The transport model BAMPS allows for the simultaneous investigation of both the suppression of high-$p_{T}$ particles and the collectivity of bulk particles consistently within a common setup that handles gluons as well as light and heavy quarks. Employing the same parameters that give a good description of the integrated elliptic flow at RHIC energies, the suppression of high-$p_{T}$ particles from gluons and light quarks is too strong. The same holds for simulations of \PbPb{} collisions at LHC, where no significant differences in the suppression pattern and the differential elliptic flow are observed compared to RHIC simulations. The integrated elliptic flow at LHC is well described up to \SI{40}{\percent} centrality, for more peripheral collisions the simulated $v_{2}$ drops too fast. Fixing the interaction of heavy quarks via elastic interactions with the medium to the suppression of heavy flavor electrons at RHIC by means of a $K$ factor, the elliptic flow of heavy flavor electrons at RHIC as well as the suppression of $D$-mesons at LHC can be reproduced.

Future studies will investigate the important effect of a running coupling on the $R_{AA}$ of gluons and light quarks and also systematically explore the implementation of the LPM effect. These modifications are qualitatively expected to bring the results for the nuclear modification factors into better agreement with experimental data. The consequences of radiative processes on the dynamics of heavy quarks will also be investigated in an upcoming study.

\section*{Acknowledgements}

This work has been supported by the Helmholtz International Center for FAIR within the framework of the LOEWE program launched by the State of Hesse. The simulations have been performed at the Center for Scientific Computing (CSC) at the Goethe University Frankfurt.

\bibliography{bibliography}

\begin{thebibliography}{10}

\bibitem{Xu:2004mz}
Zhe Xu and Carsten Greiner.
\newblock Thermalization of gluons in ultra relativistic heavy ion collisions
  by including three-body interactions in a parton cascade.
\newblock {\em Phys. Rev.}, C71:064901, 2005.

\bibitem{Gunion:1981qs}
J.~F. Gunion and G.~Bertsch.
\newblock {Hadronization by color bremsstrahlung}.
\newblock {\em Phys. Rev.}, D25:746, 1982.

\bibitem{Fochler:2010wn}
Oliver Fochler, Zhe Xu, and Carsten Greiner.
\newblock {Energy loss in a partonic transport model including bremsstrahlung
  processes}.
\newblock {\em Phys.Rev.}, C82:024907, 2010.

\bibitem{Xu:2007jv}
Zhe Xu, Carsten Greiner, and Horst Stocker.
\newblock {PQCD calculations of elliptic flow and shear viscosity at RHIC}.
\newblock {\em Phys. Rev. Lett.}, 101:082302, 2008.

\bibitem{Xu:2008av}
Zhe Xu and Carsten Greiner.
\newblock {Elliptic flow of gluon matter in ultra relativistic heavy ion
  collisions}.
\newblock {\em Phys. Rev.}, C79:014904, 2009.

\bibitem{Fochler:2008ts}
Oliver Fochler, Zhe Xu, and Carsten Greiner.
\newblock {Towards a unified understanding of jet quenching and elliptic flow
  within perturbative QCD parton transport}.
\newblock {\em Phys. Rev. Lett.}, 102:202301, 2009.

\bibitem{Adare:2008qa}
A.~Adare et~al.
\newblock {Suppression pattern of neutral pions at high transverse momentum in
  Au+Au collisions at {$\sqrt{s_{_{NN}}} = 200$} GeV and constraints on medium
  transport coefficients}.
\newblock {\em Phys. Rev. Lett.}, 101:232301, 2008.

\bibitem{Aamodt:2010jd}
K.~Aamodt et~al.
\newblock {Suppression of charged particle production at large transverse
  momentum in central Pb+Pb collisions at $\sqrt{s_{_{NN}}} = 2.76$ TeV}.
\newblock {\em Phys.Lett.}, B696:30--39, 2011.

\bibitem{Albino:2008fy}
S.~Albino, B.A. Kniehl, and G.~Kramer.
\newblock {AKK update: Improvements from new theoretical input and experimental
  data}.
\newblock {\em Nucl.Phys.}, B803:42--104, 2008.

\bibitem{Uphoff:2010sh}
Jan Uphoff, Oliver Fochler, Zhe Xu, and Carsten Greiner.
\newblock {Heavy quark production at RHIC and LHC within a partonic transport
  model}.
\newblock {\em Phys.Rev.}, C82:044906, 2010.

\bibitem{Uphoff:2010bv}
Jan Uphoff, Oliver Fochler, Zhe Xu, and Carsten Greiner.
\newblock {Heavy quarks at RHIC and LHC within a partonic transport model}.
\newblock {\em Nucl.Phys.}, A855:444--447, 2011.

\bibitem{Uphoff:2011ad}
Jan Uphoff, Oliver Fochler, Zhe Xu, and Carsten Greiner.
\newblock {Elliptic flow and energy loss of heavy quarks in ultra-relativistic
  heavy ion collisions}.
\newblock {\em Preprint} arXiv:1104.2295, 2011.

\bibitem{alice_dmeson}
Andrea~Rossi for~the ALICE~collaboration.
\newblock {D meson nuclear modification factors in Pb-Pb collisions at
  $\sqrt{s_{_{NN}}} = 2.76$ TeV, measured with the ALICE detector}.
\newblock {\em Preprint} arXiv:1106.5931, 2011.

\bibitem{Aamodt:2010pa}
K.~Aamodt et~al.
\newblock {Elliptic Flow of Charged Particles in {Pb+Pb} Collisions at
  $\sqrt{s_{_{NN}}} = 2.76$ TeV}.
\newblock {\em Phys. Rev. Lett.}, 105:252302, 2010.

\end{thebibliography}

\end{document}